\def\eqq{{\buildrel?\over=}}
\def\C{{\bf C}}

\def\R{{\bf R}}
\def\Z{{\bf Z}}
\def\Arctan{\mathop{\rm Arctan}\nolimits}
\def\Log{\mathop{\rm Log}\nolimits}
\def\Sqrt{\mathop{\rm Sqrt}\nolimits}
\def\arccosh{\mathop{\rm arccosh}\nolimits}

\documentclass[10pt, conference, compsocconf]{IEEEtran}
\usepackage{cite}
\usepackage[cmex10]{amsmath}
\usepackage{verbatim}
\usepackage{hyperref}
\usepackage{fixltx2e}
\usepackage{stfloats}

\newtheorem{lemma}{Lemma}

\newtheorem{challenge}{Challenge}
\usepackage{graphicx} 
\usepackage{url}
\begin{document}
\title {Program Verification  in the presence of complex numbers, functions with branch cuts etc.}
\author{\IEEEauthorblockN{James H. Davenport, Russell Bradford, Matthew England \& David Wilson}
\IEEEauthorblockA{Department of Computer Science\\
University of Bath\\
Bath, BA2 7AY, U.K. \\
Email: {\tt \{J.H.Davenport,R.J.Bradford,M.England,D.J.Wilson\}@bath.ac.uk}}
}
\maketitle
\begin{abstract}\noindent
In considering the reliability of numerical programs, it is normal to ``limit our study to the semantics dealing with numerical precision'' (Martel, 2005). 
On the other hand, there is a great deal of work on the reliability of programs that essentially ignores the numerics. The thesis of this paper is that there is a class of problems that fall between these two, which could be described as ``does the low-level arithmetic implement the high-level mathematics''. Many of these problems arise because mathematics, particularly the mathematics of the complex numbers, is more difficult than expected: for example the complex function log is not continuous, writing down a program to compute an inverse function is more complicated than just solving an equation,  and many algebraic simplification rules are not universally valid.
\par
The good news is that these problems are {\it theoretically\/} capable of being solved, and are {\it practically\/} close to being solved, but not yet solved, in several real-world examples. However, there is still a long way to go before implementations match the theoretical possibilities.
\end{abstract}

\section{Introduction}
It is customary, even though not often explicitly stated, to think that programming errors in numerical programs come in three distinct flavours, which we can categorise as follows.
\begin{description}
\item[1.]{\bf Blunder} (of the coding variety) This is the sort of error traditionally addressed in ``program verification'': are all array elements properly initialised before use, ``fence post'' errors\footnote{From the old puzzle ``A farmer wishes to make a 100-metre fence with supporting posts every 10 metres --- how many posts are needed'', to which the answer is 11, since each end needs a post.}, are array bounds always respected etc.? These problems are essentially independent of the numerics of the problem, and indeed are normally taught/described in integer contexts.
\item[2.]{\bf Parallelism} Issues of deadlocks or races occurring due to the parallelisation of an otherwise correct sequential program. Again, these problems are essentially independent of the numerics of the problem.
\item[3.]{\bf Numerical} Do truncation and round-off errors, individually or combined, mean that the program computes approximations to the ``true'' answers which are out of tolerance. This is the area traditionally addressed in Numerical Analysis. There are really two subquestions here: the rounding question, i.e. does $\R_{IEEE}$ (or whatever other arithmetic we are using) approximate $\R$ sufficiently well, and the truncation error question, e.g. is the discretisation $h$ small enough that it is the mathematical $\epsilon$ or is $\sum_1^N$ equivalent to $\sum_1^\infty$. Unfortunately the two interact; for example reducing $h$ in $f'(x)\approx\frac{f(x+h)-f(x)}h$ to reduce the truncation error increases the rounding problems.
\end{description}
We note that \cite{Cousot2005}, taken as a specimen of the verification literature, contains 30 papers, of which only \cite{Martel2005} deals with strictly numerical issues, four with parallelism issues, and the rest (83\%) with the first kind.
\par
It is the thesis of this paper that there is a fourth kind of error, which we can describe as follows
\begin{description}
\item[4.]{\bf Manipulation} A piece of algebra, which is ``obviously correct'', turns out not to be correct when interpreted, not as abstract algebra, but as the manipulation of functions $\R\rightarrow\R$ or $\C\rightarrow\C$.
\end{description}
{\bf Note:} throughout this paper we take the standard definitions of the branch cuts of the elementary functions from \cite{AbramowitzStegun1964,NIST2010} (as tightened in \cite{Corlessetal2000}). Other definitions would have different, but not fewer, problems. We also use the Anglo-Saxon convention that $\log$ etc. (and $\sqrt{\,}$) denote single-valued functions ($\log 1=0$, $\sqrt4=2$), whereas $\Log$ etc. denote multi-valued functions ($\Log(1)=\{2k\pi i: k\in\Z\}$, $\Sqrt(4)=\{2,-2\}$).
\par
The problems we are going to describe arise largely (though not entirely\footnote{See section \ref{sec:real} for a counter-example.
}) from complex numbers, and it is sometimes said ``real programs don't use complex numbers'', despite the fact that the introduction of \verb+COMPLEX+ into Fortran II was probably the first instance of a language data type that did not correspond to a machine data type. The authors know of several major uses of complex numbers and complex analysis, in particular many problems which arise in fluid dynamics, where two-dimensional real space $\R^2=\{(x,y)\}$ is viewed as the complex plane $\C=\{z=x+iy\}$. It is then normal to map this copy of $\C$ to another (in which the variable is traditionally denoted $w$ or $\zeta$) where the problem is easier to solve. Such an analytic map $z\mapsto w$ is termed a {\it conformal\/} map.

\section{The Kahan Example}\label{sec:Kahan}

This example comes from \cite[pp. 187--189]{Kahan1987b}, and the ultimate motivation is fluid flow in a slotted strip ($z$ space), which we wish to transform to a more convenient shape ($w$ space).
\par
\begin{figure}[h] 
\begin{center}
\includegraphics[width=3cm]{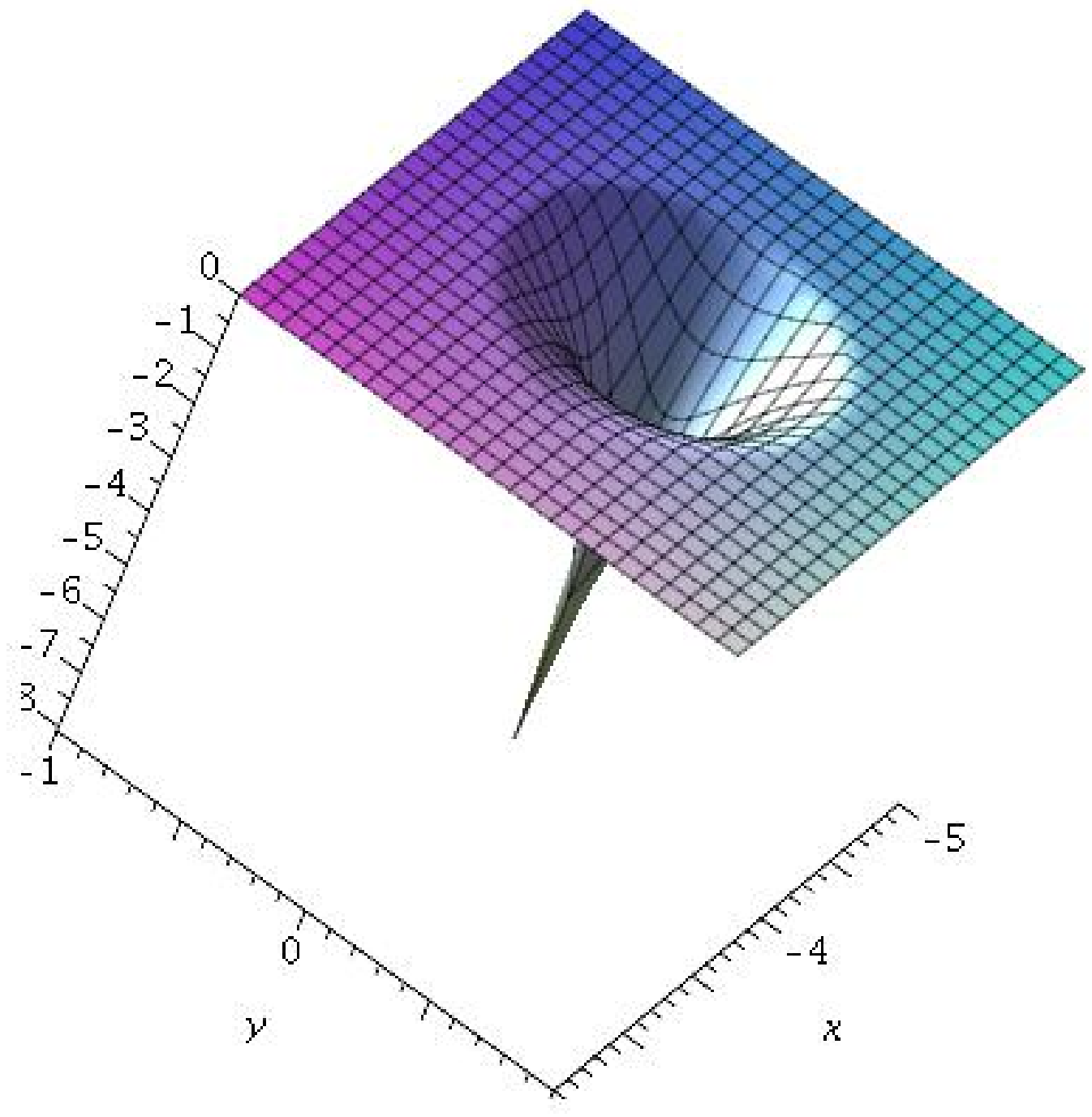}
\hspace*{0.2cm}
\includegraphics[width=3.5cm]{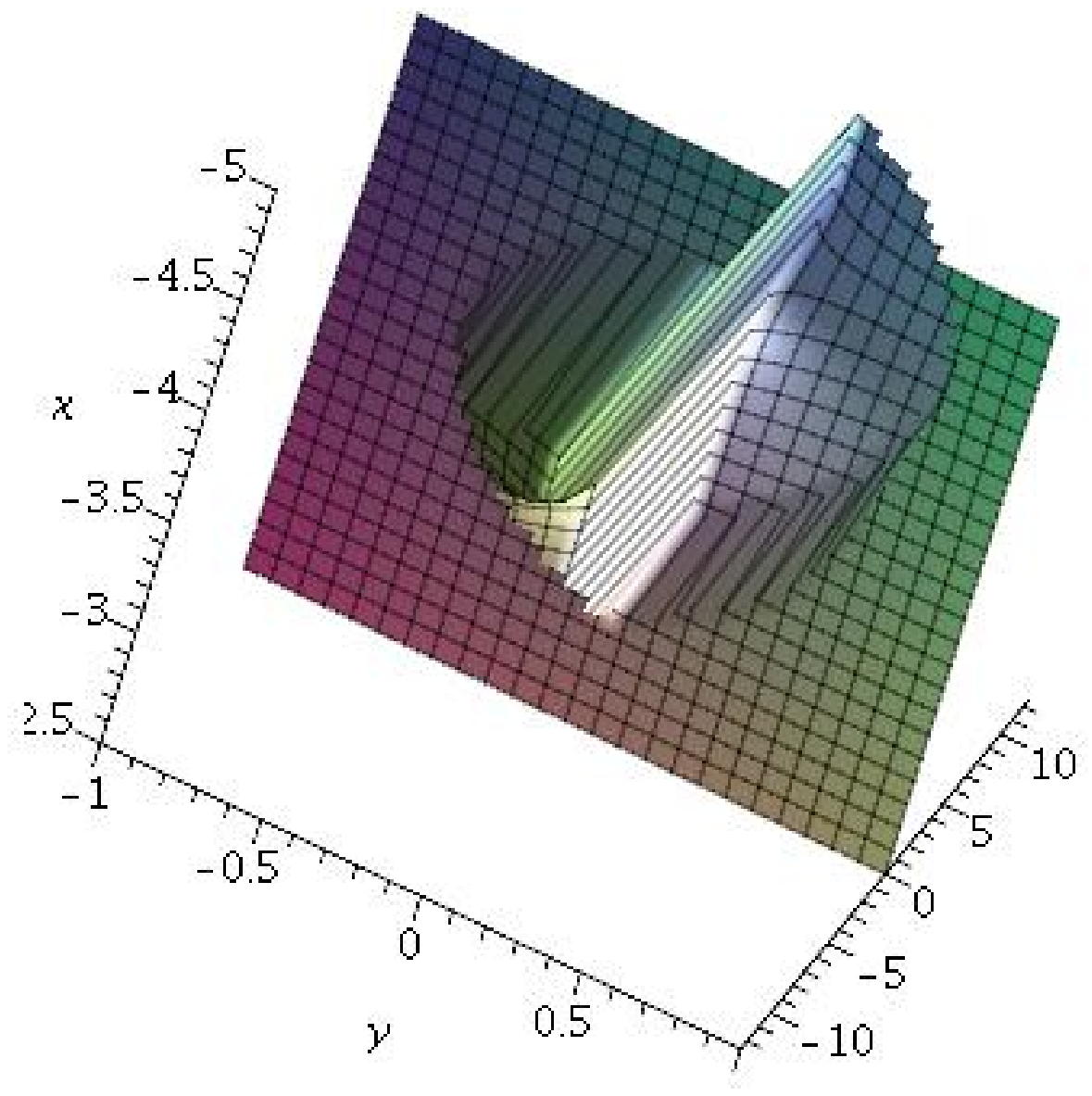}
\end{center}
\caption{Plots of the real and imaginary parts of $g(z)-q(z)$.} 
\label{fig:Kahan3D}
\end{figure}

With the usual definitions, the necessary conformal map
\begin{equation}\label{eq:K1}
w=g(z):= 2\arccosh\left(1+\frac{2z}3\right)-\arccosh\left(\frac{5z+12}{3(z+4)}\right)
\end{equation}
is only the same as the ostensibly more efficient
\begin{equation}\label{eq:K2}
w\eqq q(z):= 2\arccosh\left(2(z+3)\sqrt{\frac{z+3}{27(z+4)}}\right),
\end{equation}
if we avoid the teardrop shaped area
\begin{equation}\label{eq:Koval}
  \bigg\{z=x+iy : |y|\le \sqrt{\frac{-(x+3)^2(2x+9)}{2x+5}}, -\frac{9}{2} \le x \le -3 \bigg\}
\end{equation}
as shown by Figure \ref{fig:Kahan3D}

In fact, most computer algebra systems will refuse, these days, to convert one into the other, but this does not constitute a proof of difference.
\begin{challenge}\label{C:K1}
Demonstrate automatically that $g$ and $q$ are not equal, by producing a $z$ at which they give different results.
\end{challenge}
The technology described in \cite{Beaumontetal2007} would answer this question by decomposing the complex plane into regions, via means of cylindrical algebraic decomposition (CAD) with respect to the branch cuts of the functions, and testing the identity at a sample point in each region

\begin{figure}[h] 
\begin{center}
\includegraphics[width=5cm]{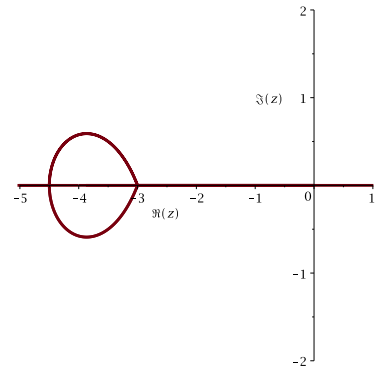}
\end{center}
\caption{Plot of the potential branch cuts for $g(z)=q(z)$ produced by the \texttt{BranchCuts} package.} 
\label{fig:KahanBC}
\end{figure}

A fully-automated implementation for this example has yet to be produced since the geometry can be quite involved.  (See section \ref{sec:Kahan} for details.)  However, progress is currently being made on the problem.  The \texttt{BranchCuts} package \cite{BranchCuts} developed at Bath and to be included with Maple 17 does isolate the curve   
\begin{equation}\label{eq:curve}
y=\pm\sqrt{\frac{(x+3)^2(-2x-9)}{2x+5}}
\end{equation}
with the appropriate $x$ range as a potential obstacle (it is the branch cut of $q$).  However, it is just one of a set of cuts returned by the code.  The plotting option in the package produces Figure \ref{fig:KahanBC} which presents the teardrop and the entire real axis as potential cuts.  

The package calculates cuts for functions by mapping the defining branch cuts of a function by the argument.  The output is generated using Maple's \texttt{solve} facilities, and the user can choose to look for solutions in the complex variable, or to first split into real and imaginary parts and work over the reals.  The first method is quicker and more widely applicable, but the second produces more intuitive output including the algebraic description of the teardrop in equation (\ref{eq:curve}).  

The package identifies the potential branch cuts of a composition of functions (a sum, a product or an equality) as the union of the cuts for individual components.  Hence the identification of the real axis as a potential obstacle is not surprising since the individual terms do have branch cuts here, with the resulting discontinuities happening to cancel out in the composition.  

\begin{figure}[h] 
\begin{center}
\includegraphics[width=5cm]{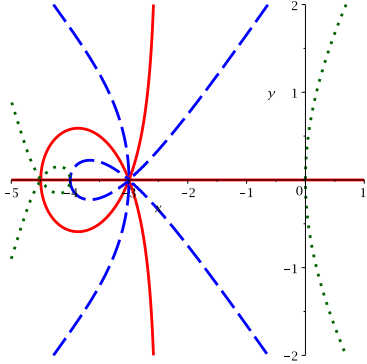}
\end{center}
\caption{Plot of the information in (\ref{eq:KahanBCSA}).  The solid line is the equality, the dashed line the first inequality and the dotted line the second.} 
\label{fig:KahanBCSA}
\end{figure}

However, the output described here would not be suitable for the technology of \cite{Beaumontetal2007} since the input into CAD must be semi-algebraic.  We can modify the code to  just return the polynomial equalities and inequalities that define each set of cuts.  For this example, there are 7 such sets, one of which consists of the three relations below.
\begin{align}
4y(2y^2x+2x^3+5y^2+21x^2+72x+81)&=0 \nonumber
\end{align}
\begin{align}
4(y^4-x^4+3y^2x-13x^3 \qquad \,\, & \nonumber \\
+9y^2-63x^2-135x-108)&<0 \nonumber \\
4x^4-4y^4-12y^2x+52x^3 \qquad \nonumber \\
-63y^2+225x^2+324x&<0  \label{eq:KahanBCSA}
\end{align}
Figure \ref{fig:KahanBCSA} gives a plot of these three curves.  By testing sample points we see where the inequalities are satisfied and infer that the branch cut defined is the teardrop along with the real axis above $-3$.  These issues and the implementation of the \texttt{BranchCuts} package are discussed further in \cite{BranchCuts}.

\begin{figure}[h] 
\begin{center}
\includegraphics[width=3.5cm]{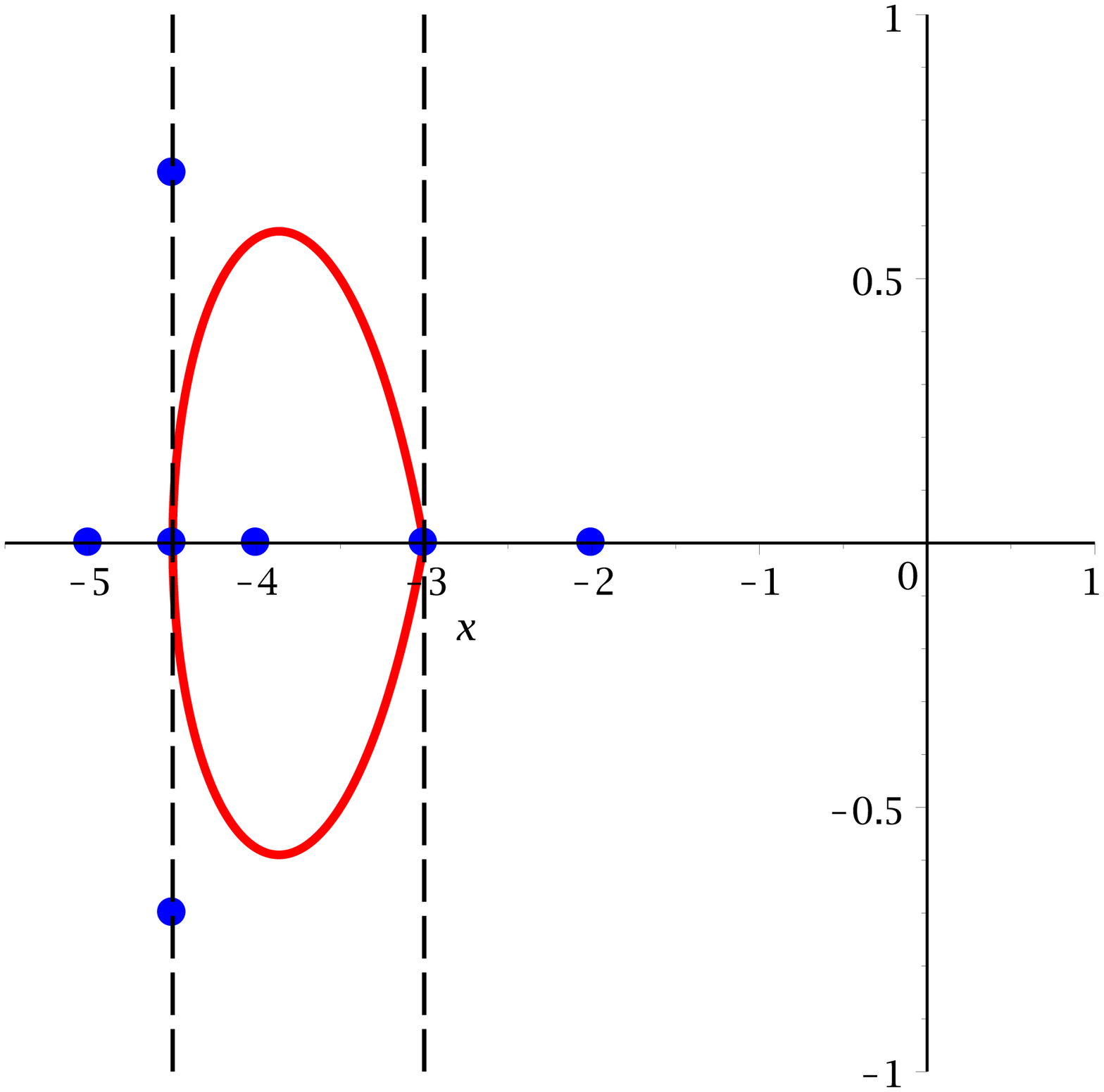}
\hspace*{0.2cm}
\includegraphics[width=3.5cm]{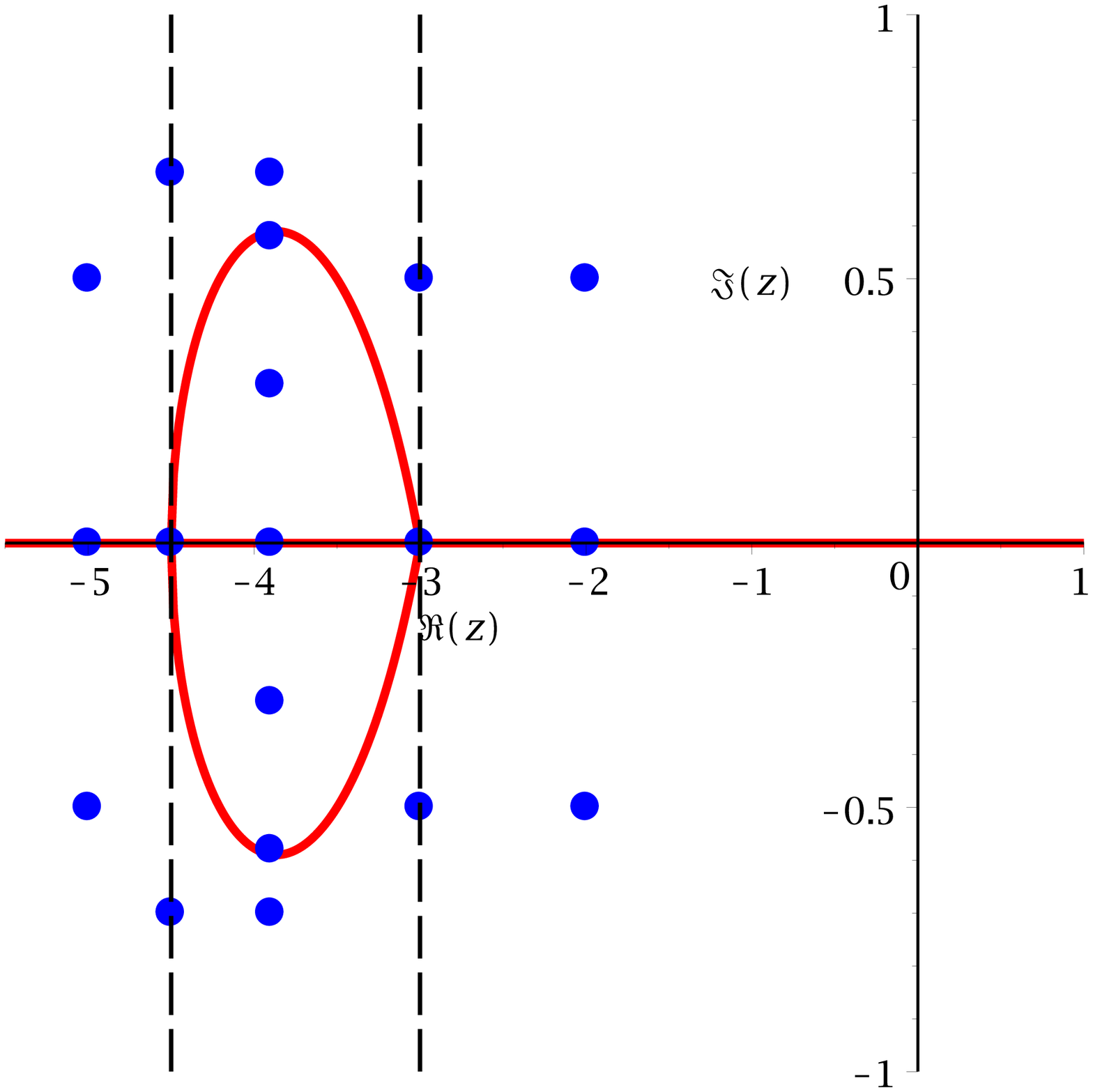}
\end{center}
\caption{Possible minimal CADs for Kahan's example.  The dots indicate sample points for a cell.} 
\label{fig:KahanMinCAD}
\end{figure}

If we pass the full set of polynomials to CAD (ignoring whether they are equalities or inequalities) then clearly a lot more information will be processed than required.  Using the variable ordering $y>x$ and the command \texttt{CylindricalAlgebraicDecompose} within Maple 16 \cite{Chenetal2009d}, this produces a CAD of 409 cells.  Given (\ref{eq:Koval}) we might hope for a minimal CAD of 13 cells, or if we accept that the real axis must be included in any calculations then a minimal CAD of 19 cells, (see Figure \ref{fig:KahanMinCAD}).  

Note that there is a simplification of $g$ {\it valid over the whole complex plane \/}: $g$ can legitimately be rewritten to
\begin{equation}\label{eq:K3}
w=h(z):=
2\,\ln  \left( \frac13\,{\frac {\sqrt {3\,z+12} \left( \sqrt {z+3}+\sqrt {
z} \right) ^{2}}{2\,\sqrt {z+3}+\sqrt {z}}} \right) ,
\end{equation}
The  technology in \cite{Beaumontetal2007}  can show this, i.e. $\forall z\in\C \, g(z)=h(z)$, but there are multiple square roots requiring denesting \cite[\S4.3]{Phisanbut2011a} and (as formulated) square roots in the denominator.
Indeed the standard tools of Maple 16 currently get this wrong: \verb+coulditbe(g<>h);+ returns \verb+true+, which {\it ought\/} to indicate that there is a counter-example.
\begin{challenge}\label{C:K2}
Demonstrate automatically that $g$ and $h$ are equal.
\end{challenge}

Although the technology in \cite{Beaumontetal2007}, implemented in a mixture of Maple and QEPCAD (though we are working on a purely Maple implementation based on \cite{Chenetal2009d}), may be able to meet this challenge in time, we would be left with the problem of trusting the underlying demonstration code.  So there is the additional problem of translating this methodology into a tool such as MetiTarski \cite{Paulson2012b}.

\section{Joukowski Examples}
Consider the Joukowski map \cite[pp. 294--298]{Henrici1974}: 
\begin{equation}\label{eq:J}
f:z\mapsto \frac12\left(z+\frac1z\right).
\end{equation}
\subsection{Joukowski (a)}
\begin{lemma}\label{l:1}
$f$ is injective as a function from $D:=\{z:|z|>1\}$.
\end{lemma}
If $z\mapsto \zeta$ then $1/z\mapsto \zeta$, and there are no other pre-images of $\zeta$ (since the algebraic inverse of (\ref{eq:J}) is the solution of a quadratic). If $|z|>1$, then $|1/z|<1$, so $z$ is unique in $D$.
\par
In fact $f$ is a bijection from $D$ to $\C^\ddag:=\C\setminus[-1,1]$, and hence has an inverse.
\par
Of course, (\ref{eq:J}) is the conformal map $\C\rightarrow\C$ that equates to the map
\begin{equation}\label{eq:JR}
f_R:(x,y)\mapsto \left(
\frac12\,x+\frac12\,{\frac {x}{{x}^{2}+{y}^{2}}},
\frac12\,y-\frac12\,{\frac {y}{{x}^{2}+{y}^{2}}} \right)
\end{equation}
$\R^2\rightarrow\R^2$. However, it is not obvious from  (\ref{eq:JR}) alone that $f_R$ is a bijection $D\rightarrow\C^\ddag$, i.e. that
\begin{equation}\label{eq:JRbi}
\begin{array}{cc}\forall x_1\forall x_2\forall y_1\forall y_2&
\biggl(x_1^2+y_1^2>1\land x_2^2+y_2^2>1 \land\\
&x_1+\,{\frac {x_1}{x_1^{2}+y_1^{2}}}=
\,x_2+\,{\frac {x_2}{x_2^{2}+y_2^{2}}}\land\\&
\,y_1-\,{\frac {y_1}{x_1^{2}+y_1^{2}}} =
\,y_2-\,{\frac {y_2}{x_2^{2}+y_2^{2}}}\biggr) \Rightarrow\\& \bigl(x_1=x_2\land y_1=y_2\bigr).
\end{array}
\end{equation}
\begin{challenge}\label{C:J1}
Demonstrate automatically the truth of (\ref{eq:JRbi}), which is also \cite[(49)]{Wilson2012a}.
\end{challenge}
We have been unable to do this with either the QEPCAD \cite{Brown2003} implementation of Partial Cylindrical Algebraic Decomposition \cite{CollinsHong1991} or the Maple implementation of  Cylindrical Algebraic Decomposition via triangular decomposition \cite{Chenetal2009d}.
\par
Brown \cite{Brown2012a} suggested writing problems of the form $\forall {\bf v} A\rightarrow(f=0\land g=0)$ as $\lnot \exists  {\bf v} A\land\lnot(f=0\land g=0)$ . This splits into two clauses: \cite[\S7.2]{Wilson2012a}, which can be seen to have equational constraints \cite{McCallum1999a}.
\par
Even these are beyond QEPCAD and Maple currently.
However, Brown \cite{Brown2012a} has been able to recast these clauses (manually) to make them amenable to QEPCAD, and indeed solved the problem in under 12 seconds.
\begin{challenge}\label{C:J1-auto}
Automate these techniques and transforms.
\end{challenge}
Having established (or not) that $f$ is a bijection $D\rightarrow\C^\ddag$, we want its inverse. Formally, this is trivial, as one referee said
\begin{quote}The inverse of Joukowski is the solution of a quadratic with the usual sign
ambiguity:
\end{quote}
if $\zeta=\frac12\left(z+\frac1z\right)$, then $2z\zeta=z^2+1$ and $z=\zeta\pm\sqrt{\zeta^2-1}$. This is easily within the grasp of computer algebra, as seen in Figure \ref{fig:MJ}.
\begin{figure}
\caption{Maple's {\tt solve} on inverting Joukowski\label{fig:MJ}}
\begin{verbatim}
> [solve(zeta = 1/2*(z+1/z), z)];
\end{verbatim}
$$
\left[\zeta+\sqrt {{\zeta}^{2}-1},\zeta-\sqrt {{\zeta}^{2}-1}\right]
$$
\end{figure}
The only challenge might be the choice implicit in the $\pm$ symbol: which do we choose?
\par
Unfortunately, the answer is ``neither'', or at least ``neither, uniformly''. 
The true answer is that, for $f$ a bijection from $\{z:|z|>1\}$ to $\C\setminus[-1,1]$, its inverse is
\begin{equation}\label{eq:Ji}
f_1(\zeta)= \zeta \left\{\begin{array}{cl}
+\sqrt{\zeta^2-1} & \Im(\zeta)>0 \\
-\sqrt{\zeta^2-1} & \Im\zeta)<0 \\
+\sqrt{\zeta^2-1} & \Im(\zeta)=0 \land\Re(\zeta)>1\\ 
-\sqrt{\zeta^2-1} & \Im(\zeta)=0 \land\Re(\zeta)<-1 
\end{array}\right.
\end{equation}
In fact, a better (at least, free of case distinctions) definition is
\begin{equation}\label{eq:Ji2}
f_2(\zeta)= \zeta+\sqrt{\zeta-1}\sqrt{\zeta+1}.
\end{equation}
The techniques of \cite{Beaumontetal2007} are able to {\bf verify}  (\ref{eq:Ji2}), in the sense of showing that $f_2(f(z))-z$ is the zero function on $\{z:|z|>1\}$.
\begin{challenge}\label{C:J1bis}
Derive automatically, and demonstrate the validity of, either  (\ref{eq:Ji}) or  (\ref{eq:Ji2}). In terms of Maple, we would want to see Figure  \ref{fig:MJinvgood}, rather than the actual Figure  \ref{fig:MJinvbad}.
\end{challenge}
\begin{figure}
\caption{Maple's actual {\tt solve} on inverting injective Joukowski\label{fig:MJinvbad}}
\begin{verbatim}
> [solve(zeta = 1/2*(z+1/z), z)]\
   assuming abs(z) > 1
\end{verbatim}
$$
\left[\zeta+\sqrt {{\zeta}^{2}-1},\zeta-\sqrt {{\zeta}^{2}-1}\right]
$$
\end{figure}
\begin{figure}
\caption{Ideal Maple {\tt solve} on inverting injective Joukowski\label{fig:MJinvgood}}
\begin{verbatim}
> solve(zeta = 1/2*(z+1/z), z)\
   assuming abs(z) > 1
\end{verbatim}
$$
\zeta+\sqrt{\zeta-1}\sqrt{\zeta+1}
$$
\end{figure}
In terms of derivation, the techniques of \cite{CorlessJeffrey1996} seem worthy of investigation, but the first author has been unable to do this derivation satisfactorily by this route.
\subsection{Joukowski (b)}
Here the function is again given by  (\ref{eq:J}). 
\begin{lemma}\label{l:2}
$f$ is injective as a function from $H:=\{z:\Im z>0\}$.
\end{lemma}
As in Lemma \ref{l:1}, if $z\mapsto \zeta$ then $1/z\mapsto \zeta$, and there are no other pre-images of $\zeta$. If $\Im(z)>0$, $\Im(1/z)<0$, and $f$ in therefore injective from $H$.
\par
In fact, $f$ is  a bijection from $H$ to $\C\setminus((-\infty,-1]\cup[1,\infty))$, and hence has an inverse.
\par
Again, it is not obvious from  (\ref{eq:JR}) alone that $f_R$ is a bijection, now from $\{(x,y) | y>0\}$, i.e. that
\begin{equation}\label{eq:JRbi-ii}
\begin{array}{cc}\forall x_1\forall x_2\forall y_1\forall y_2&
\biggl(y_1>0\land y_2>0 \land\\&
\,x_1+\,{\frac {x_1}{x_1^{2}+y_1^{2}}}=
\,x_2+\,{\frac {x_2}{x_2^{2}+y_2^{2}}}\land\\&
\,y_1-\,{\frac {y_1}{x_1^{2}+y_1^{2}}} =
\,y_2-\,{\frac {y_2}{x_2^{2}+y_2^{2}}}\biggr) \\&\Rightarrow \bigl(x_1=x_2\land y_1=y_2\bigr).
\end{array}
\end{equation}
\begin{challenge}\label{C:J2}
Demonstrate automatically the truth of (\ref{eq:JRbi-ii}).
\end{challenge}
It is likely that the ideas of \cite{Brown2012a} can do this, but again these need automation.
\par
We have the same challenge over the inverse of $f$: again formally it is $f^{-1}\eqq\zeta\pm\sqrt{\zeta^2-1}$, and the only challenge is the $\pm$ symbol: which do we choose? Here \cite[(5.1-13), p. 298]{Henrici1974} argues for
\begin{equation}\label{eq:Ji2b}
f_3(\zeta)= \zeta+\underbrace{\sqrt{\zeta-1}}_{\arg\in(-\pi/2,\pi/2]}\underbrace{\sqrt{\zeta+1}}_{\arg\in(0,\pi]}.
\end{equation}
\begin{challenge}\label{C:fn}
Find a way to represent functions such as $\underbrace{\sqrt{\zeta+1}}_{\arg\in(0,\pi]}$
\end{challenge}
Fortunately this one is soluble in this case\footnote{And is probably soluble more generally, but the authors know of no general work on ``alternative formulations''.}, we can write $\underbrace{\sqrt{\zeta+1}}_{\arg\in(0,\pi]}=i
 \underbrace{\sqrt{-\zeta-1}}_{\arg\in(-\pi/2,\pi/2]}$, and the latter is the normal \verb+sqrt+ function of \cite{AbramowitzStegun1964}. Hence we have an inverse function
\begin{equation}\label{eq:Ji3}
f_4(\zeta)= \zeta+\sqrt{\zeta-1}i\sqrt{-\zeta-1}.
\end{equation}
\begin{challenge}\label{C:J2bis}
Demonstrate automatically that this is an inverse to $f$ on $\{z:\Im z>0\}$.
\end{challenge}
There is also the problem of deducing (\ref{eq:Ji3}). One could try automatic deduction on the lines of \cite{CorlessJeffrey1996}, but there is another possibility: heuristic search followed by verification {\it if\/} the verification were feasible \cite{Gulwani2012a}. 
\section{A Real Example}\label{sec:real}
\def\foo{\cite[(4.4.34)]{AbramowitzStegun1964}\cite[(4.24.15)]{NIST2010}}
Just in case the reader thinks that the real numbers are immune from these problems, consider the addition rule for the inverse tangent, quoted as (\foo)
$$
\Arctan(x)\pm\Arctan(y)=\Arctan\left(\frac{x\pm y}{1\mp xy}\right).
$$
Despite the caveat in \cite{NIST2010} that ``The above equations are interpreted in the sense that every value of the
left-hand side is a value of the right-hand side and vice versa'', it is in fact the case that the `obvious' two equations are true separately, {\tt viz.\/}
\begin{eqnarray}
\Arctan(x)+\Arctan(y)&=&\Arctan\left(\frac{x+ y}{1- xy}\right)\label{eq:Arctanplus}\\
\Arctan(x)-\Arctan(y)&=&\Arctan\left(\frac{x- y}{1+ xy}\right)
\end{eqnarray}
Consider (\ref{eq:Arctanplus}):
This is valid for the multi-valued $\Arctan$, but for the single-valued $\arctan$ only when $|1- xy|<1$, due to a ``branch cut at infinity'' of $\arctan$. Nevertheless, the single-valued version of  (\ref{eq:Arctanplus}) is often cited as true: see for example
\cite[(5.2.5)]{Terr2012a}.
\par
Over the reals, this is a non-challenge, the techniques of \cite{Beaumontetal2007} do solve it easily, and produce a counterexample.
\section{So why are these challenges?}
\subsection{Complex functions and branch cuts}
These are difficult subjects, which have tended to be brushed under the carpet. The first truly algorithmic approach is ten years old  (\cite{Bradfordetal2002}, refined in \cite{Beaumontetal2007}), and has various difficulties.
\begin{enumerate}
\item At its core is the use of Cylindrical Algebraic Decomposition of $\R^{N}$ to find the connected components of $\C^{N/2}\setminus\{\hbox{\rm branch cuts}\}$. The complexity of this is doubly exponential in $N$: upper bound of $d^{O(2^N)}$ \cite{Hong1991} and lower bounds of $2^{2^{(N-1)/3}}$ \cite{BrownDavenport2007,DavenportHeintz1988}. While better algorithms for the connected components problem are in principle known (\cite{Basuetal2012b} is $d^{O(N\sqrt N)}$), we do not know of any accessible implementations. 
\par
Furthermore, we are clearly limited to small values of $N$, at which point looking at $O(\ldots)$ complexity is of limited use. We note that the cross-over point between $2^{(N-1)/3}$ and $N\sqrt N$ is at $N=21$. A more detailed comparison is given in \cite{Hong1991}. Hence there is a need for practical research on low-$N$ Cylindrical Algebraic Decomposition.
\item While the fundamental branch cut of $\log$  is simple enough, being $\{z=x+iy | y=0 \land x<0\}$, actual branch cuts are messier. Part of the branch cut of (\ref{eq:K2}) is 
\begin{eqnarray*}
&2x^3+21x^2+72x+2xy^2+5y^2+81=0 \land\\& \hbox{other conditions},
\end{eqnarray*}
whose solution accounts for the curious expression in (\ref{eq:Koval}). While there has been some progress in manipulating such images of half-lines (described in \cite{Phisanbutetal2010a,Phisanbut2011a}), there is almost certainly  more to be done.
\end{enumerate}
\subsection{Injectivity}
Lemmas \ref{l:1} and \ref{l:2} might seem to be statements about complex functions of one variable, so why do we need to handle (or fail to handle) statements about four real variables to prove them? There are three, rather distinct, reasons for this.
\begin{enumerate}
\item The statements require the $|\cdot|$ function (Lemma \ref{l:1}) or the $\Im$  function (Lemma \ref{l:2}), neither of  which are complex analytic functions. Hence some recourse to real analysis (and therefore twice as many variables) seems inevitable, though it would be nice to have a more formal statement and proof of this.
\item Equations (\ref{eq:JRbi}) and (\ref{eq:JRbi-ii}) are the direct translations of the basic definition of injectivity. In practice, certainly if we were looking at functions $\R\rightarrow\R$, we would want to use the fact that the function concerned was continuous.
\begin{challenge}\label{C:inj}
Find a better formulation of injectivity questions $\R^N\rightarrow\R^N$, making use of the properties of the functions concerned (certainly continuity, possibly rationality).
\end{challenge}
\item While equations (\ref{eq:JRbi}) and (\ref{eq:JRbi-ii}) are statements from the existential theory of the reals, and so the theoretically more efficient algorithms quoted in \cite{Hong1991} are in principle applicable, the more modern developments described in \cite{PassmoreJackson2009} do not seem to be directly applicable. However, we can transform then into a disjunction of statements to each of which the Weak Positivstellensatz \cite[Theorem 1]{PassmoreJackson2009} is applicable.
\begin{challenge}\label{C:PJ}
Solve these problems using the techniques of \cite{PassmoreJackson2009},
\end{challenge}\end{enumerate}
\section{Conclusions}
The aim of this paper has been to demonstrate that translating mathematical problems into programs may require some algebraic manipulations whose accuracy is not as obvious as one might think, and whose verification is {\it currently\/} not as straightforward as we would like, despite the fact that their correctness is, in principle, decidable. A summary is given in Table \ref{tab:challenges}.
\par\noindent
These are, largely, concrete challenges that, we hope, will spur practical advances in this domain.
\section*{Acknowledgement}
The authors would like to thank Acyr Locatelli, Gregory Sankaran and Nicolai Vorobjov of the Bath Triangular Sets seminar, as well as Scott McCallum (Macquarie U.) who kindly visited us, for their input, Chris Brown (U.S.~Naval Academy) for \cite{Brown2012a}, and the referees for their comments, but the errors and omissions are all the authors.
\newpage
\begin{table}[h]
\caption{Current state of these challenges\label{tab:challenges}}
\begin{tabular}{ll}
Challenge&State\cr
\ref{C:K1}&Mathematically solved by \cite{Beaumontetal2007}, \\ & geometry taxes current solvers.\cr
\ref{C:K2}&Mathematically solved by \cite{Beaumontetal2007}, \\ & branch cut determination\cr&not yet implemented.\cr
\ref{C:J1}/\ref{C:J2}&Mathematically solved by \cite[etc.]{Collins1975}, \\ & geometry defeats current solvers.\cr
\ref{C:J1-auto}&Under development.\cr
\ref{C:J1bis}/\ref{C:J2bis}& Mathematically solved \cite{Beaumontetal2007}, geometry defeats current \cr& solvers, and is probably significantly harder than \\
 & in the previous problems.\cr
\ref{C:fn}&Unknown: probably straightforward research project.\cr
\ref{C:inj}&Unknown: research project.\\
\ref{C:PJ}&Unknown: project for the authors of \cite{PassmoreJackson2009}.\cr
\end{tabular}
\end{table}
\hyphenpenalty=100
\bibliography{../../../../../jhd}
\end{document}